\documentclass[prb,twocolumn,showpacs,preprintnumbers,amsmath,amssymb]{revtex4-1}
\usepackage{graphicx}

\begin{document}
\title{Quantum Lifshitz transitions generated by order from quantum disorder
in strongly correlated Rashba spin-orbital coupled systems }
\author{ Fadi Sun$^{1,2}$ and Jinwu Ye $^{1,2,3}$   }
\affiliation{
$^{1}$ Tsung-Dao Lee Institute, Shanghai 200240, China \\
$^{2}$ Institute for Quantum Science and Engineering, Shenzhen 518055, China   \\
$^{3}$ Department of Physics and Astronomy, Mississippi State University, MS, 39762, USA  }
\date{\today }

\begin{abstract}
%Inspired by recent experimental advances to generate 2d  Rashba spin-orbit coupling (SOC) for cold atoms,
We study the system of strongly interacting spinor bosons in a square lattice subject to
the isotropic Rashba SOC $ \alpha=\beta $. It supports collinear spin-bond correlated magnetic Y-x phase, 
a gapped in-commensurate (IC-) co-planar IC-XY-y phase, a non-coplanar commensurate (C-) $ 3 \times 3 $ Skyrmion crystal phase (SkX).
The state at the Abelian point $ \alpha=\beta=\pi/2 $ is just an AFM state in a rotated basis.
Slightly away from the point, we identify a spurious $ U(1) $ symmetry,
develop a novel and non-perturbative method to calculate not only the gap, but also the excitation spectrum
due to the order from quantum disorder (OFQD) mechanism. We construct a symmetry based effective action to investigate the quantum Lifshitz
transition from the Y-x state to the IC-XY-y state and establish the connection between
the phenomenological  parameters in the effective action and those evaluated by the microscopic non-perturbative OFQD analysis in the large $ S $ limits.
Experimental implications on cold atoms and some 4d or 5d Kitaev materials are discussed.
\end{abstract}

\maketitle

{\sl 1. Introduction. }
 It was well known that geometric frustrations lead to fantastic quantum, topological phases
 and phase transitions in quantum spin systems \cite{sachdev,aue,SLrev1,SLrev3}.
% The strong atomic spin-orbit coupling in some 4d Ruthenium \cite{cl1,cl2} or 5d Iridium
% \cite{kitaevlattice,kitaevlattice1,kitaevlattice2} materials may lead to
% bond dependent spin exchange interactions in various quantum compass models \cite{SLrev1,SLrev3}.
 Novel frustrated phenomena in some typical quantum compass models
 such as the Kitaev honeycome lattice model \cite{kit}, $120^{\circ} $ honeycomb lattice model \cite{pband1,pband2,tripod},
 and Heisenberg-Kitaev model \cite{kit123} have also been studied.
 On the other forefront, Rashba spin-orbit coupling (SOC) is ubiquitous in various 2d or layered
 non-centrosymmetric magnetic insulators, semi-conductor systems, metals and superconductors \cite{rashba,ahe,socsemi,ahe2,she,niu,aherev,sherev}.
 There were also experimental advances in generating various kinds of 2D SOC for charge neutral cold atoms in
 both continuum and optical lattices \cite{expk40,expk40zeeman,2dsocbec,3dsocweyl}.
 New experimental schemes \cite{clock,clock1,clock2,SDRb,ben} were successfully implemented to create
 a long-lived SOC gas of quantum degenerate atoms.
 These cold atom experiments set-up a very promising platform to observe  many-body phenomena  due to
 the interplay between Rashba SOC and interaction in optical lattices.
% There are also recent experimental advances in generating 2d Rashba  SOC for
% both fermions and spinor bosons in optical lattices \cite{expk40,expk40zeeman,2dsocbec,clock,clock1,clock2,SDRb,ben,sorev}.
% New many body phenomena due to the interplay among strong interactions,
% the SOC and lattice geometries  are being investigated in the current cold atom experiments.
 It becomes important to investigate what would be the new quantum or topological phenomena due to such an interplay.
% the interplay between the strong correlations and the Rashba SOC on lattices.

In this work, we address this outstanding problem by studying the system of strongly interacting spinor bosons in a square lattice subject to
the 2d Rashba SOC. We find  that the Rashba SOC provides a new class of frustrated source
%even in a square lattice
%different from the well known geometric frustrations and the quantum compass models.
which leads to novel and rich quantum phenomena even in a square lattice summarized in the abstract and Fig.1.
%They include masses generated from "order from quantum disorder ", the collinear spin-bond correlated  Y-x phase
%and its novel excitation spectrum,
%quantum commensurate ( C ), especially the In-commensurate ( IC ) non-coplanar Skyrmion crystal ( SkX ) phases,
%stripe co-planar ( spiral ) C and  quantum Lifshitz C-IC transitions, etc.
Our results can be applied to ongoing and near future cold atom experiments as soon as the heating issues can be overcame
in the strong coupling limit.
%and also non-centrosymmetric magnetic insulators.
They may also shed considerable lights on the un-conventional magnetic ordered states or
putative quantum spin liquid states in some 4d or 5d Kitaev materials \cite{SLrev1,SLrev3}.

%Our results demonstrate that the Rashba SOC induced frustration
%opens a new avenue to explore whole new classes of quantum or topological phenomena which
%may have wide implications in cold atoms and various materials with SOC to be discussed near to the end of the paper.
%Some possible far-reaching perspectives are given.

The tight-binding Hamiltonian of ( pseudo)-spin $ 1/2 $ bosons ( fermions )  hopping in
a two-dimensional square lattice subject to any combination of Rashba and Dresselhaus SOC is\cite{wu,classdm1,classdm2,rh}:
\begin{equation}
	\mathcal{H}_{B}= -t\sum_{\langle ij\rangle}(b_{i\sigma}^\dagger U_{ij}^{\sigma\sigma'} b_{j\sigma'}+h.c.)
	+ \frac{U}{2} \sum_i ( n_{i}-n )^2
\label{hubbardint}
\end{equation}
where $ t $ is the hopping amplitude along the nearest neighbors $\langle ij\rangle$, $ n $ is taken to be an integer filling,
$U_{i,i+\hat{x} } =e^{i \alpha \sigma_x}$, $U_{i.i+\hat{y} }=e^{i \beta \sigma_y}$ are the non-Abelian gauge fields
 put on the two links in a square lattice. $U>0$ is the Hubbard onsite interaction.

In the strong coupling limit $ U/t \gg 1 $,  to the order $O(t^2/U)$,
we obtain the effective spin $ s=n/2 $ Rotated Ferromagnetic  Heisenberg model (RFHM) \cite{rh}:
\begin{equation}
	\mathcal{H}_{R}  =  -J\sum_i
	[\mathbf{S}_i R(\hat{x},2\alpha)\mathbf{S}_{i+\hat{x}}
	+\mathbf{S}_i R(\hat{y},2\beta)\mathbf{S}_{i+\hat{y}}]
\label{rhgeneral}
\end{equation}
with $J= \pm 4t^2/U > 0 $ for bosons/fermions,
the $R(\hat{x},2\alpha)$, $R(\hat{y},2\beta)$ are the two SO(3) rotation matrices
around the $ X $ and $ Y $ spin axis by angle $2\alpha$, $2\beta$
putting on the two bonds along  $\hat{x} $, $\hat{y} $ respectively.
Expanding  $ U_{i,i+\hat{x}}= \cos \alpha + i \sin \alpha \sigma_x,
U_{i,i+\hat{y}}= \cos \beta + i \sin \beta \sigma_y $ in Eq.1, one can see that at  the Abelian point $ \alpha=\beta=\pi/2 $,
the standard hopping terms vanish, only the spin-flip hopping term ( SOC) survive.
As shown in \cite{rh}, at the Abelian point, Eq.2 is simply the FM Heisenberg model in the rotated
$ \tilde{\tilde{SU}}(2) $  basis $ H= -J \sum_{ ij } \vec{\tilde{\tilde{S}}}_{i} \cdot \vec{\tilde{\tilde{S}}}_{j} $
where $ \vec{\tilde{\tilde{S}}}_{i} = R(\hat{x},\pi n_1)  R(\hat{y},\pi n_2) \vec{S}_{i} $.
%So there is a hidden $ SU(2) $ symmetry at the Abelian point.

%The fermions will be discussed in a separate publication.

%To understand quantum phenomena in the generic SOC parameter space of Eqn.\ref{rhgeneral}, we take a ``divide and conquer'' strategy.
%We first explore new and rich quantum phenomena along the
%solvable extremely anisotropic limit $\alpha=\pi/2, 0<\beta<\pi/2$.
%Then starting from the deep knowledge along the anisotropic  line,
%we will try to investigate the quantum phenomena at generic $(\alpha,\beta)$.
%The first step has been achieved in \cite{rh}, here, we will achieve the second step.
%This strategy has been successful in solving several strongly coupled models such as Kondo model \cite{kondo0,kondo1,kondo2}
%and quantum dimer model \cite{dimer1,dimer2}.

Both Eq.\ref{hubbardint} and Eq.\ref{rhgeneral} at a generic $ (\alpha, \beta ) $ have the translational,
the time reversal $ {\cal T} $, the three spin-orbital coupled $ Z_2 $
symmetries $ {\cal P}_x, {\cal P}_y, {\cal P}_z $ symmetries  \cite{rh}.
Along the isotropic Rashba limit $ \alpha=\beta $, the $ {\cal P}_z $ symmetry is enlarged to the spin-orbital coupled $ [C_4 \times C_4]_D $ symmetry around the $ z $ axis. In this paper, we focus on spinor bosons with the isotropic Rashba SOC $ \alpha= \beta $.
The generic case $ \alpha \neq \beta $ is presented in a separate publication \cite{unlong}.

% Some applications of the RFHM to these materials have been discussed in \cite{rhh} and will be discussed further near to the end of this paper.

\begin{figure}[!htb]
	%\centering
\includegraphics[width=0.45\textwidth]{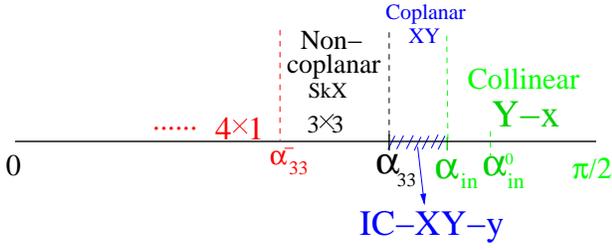}			
\caption{  The phase diagram of Eq.2 when $  \alpha= \beta^{+} $.
 The state at $ \alpha=\beta=\pi/2 $ is just an AFM state in $ \tilde{\tilde{SU}}(2) $ basis.
% There is a co-planar IC-XY-y phase intervening between the collinear Y-x phase and the non-Co-planar SkX phase
 When $  \alpha_{in} < \alpha < \pi/2 $, there is a	 gap opening in the collinear Y-x phase
 generated by the order from quantum disorder (OFQD) mechanism.
% The gap keeps increasing  when moving away from the Abelian point $ \alpha=\beta=\pi/2 $.
% the Y-x state becomes metastable between the first order transition
% ( diagonal ) line $  \alpha_{in} < \alpha < \pi/2 $ and the second order transition ( dashed ) line
% driven by the condensations of the $ C-C_{\pi} $ and C-IC with $ \pi- q_{ic} < k^{0}_y < \pi $.
 There is a second order quantum Lifshitz transition (QLT) at $ \alpha= \alpha_{in} $ with the dynamic exponent $ z=1 $,
 from the Y-x phase to the coplanar IC-XY-y phase \cite{notation}, then a second one to the C- non-coplanar
 $ 3 \times 3 $ SkX  phase at $ \alpha= \alpha_{33} $.
% The small domes stand for all the intervening phases between any two gapped principle $ N \times 1, N \geq 3$ phases:
% the C- spiral gapped phases at higher order $ \alpha =n/N \pi, n>1 $ with much smaller widths and also the gapless
% IC-YZ-x phases \cite{notation} taking zero measures and forming a Cantor set with a fractal dimension \cite{coexist}.
% It is a bi-critical point  where
% the two second order transition lines from the $ 3 \times 3 $ SkX to the IC-SkX/Y-x below the diagonal line
% and to the IC-SkX/X-y above the diagonal line meet the first order transition line between the IC-SkX/Y-x and IC-SkX/X-y along the diagonal line.
% The  $ 3 \times 3 $ SkX  and IC-SkX is the only non-coplanar C and IC phase along the diagonal line.
% The quantum order from disorder phenomenon at $  \alpha_{in} < \alpha < \pi/2 $.
% The state at $ \alpha=\beta=\pi/2 $ is just an AFM state in $ \tilde{\tilde{SU}}(2) $ basis \cite{hidden}.
% The state near $ \alpha=\beta=0 $ is some IC-YZ-x phase instead of a FM  shown in the SM \cite{never}.
 The relevant numbers are $ \alpha^{0}_{in} \sim 0.3611 \pi, \alpha_{in} \sim 0.3526 \pi,
 \alpha_{33} \sim 0.3402 \pi,  \alpha^{-}_{33} \sim 0.295 \pi$ and the ordering
 wavevector in the IC-XY-y is $ \pi-q^{0}_y $ with  $ q_{ic} \sim 0.18 \pi < q^{0}_y < 0.24 \pi $.
 When $  \alpha= \beta^{-} $. All the phases become their corresponding imaging phases related by the $ [C_4 \times C_4]_D $ transformation
 except the $ 3 \times 3 $ SkX  phase is its own image.  So the two corresponding imaging phases can coexist with any ratio
 along $ \alpha=\beta $. }
\label{phasedia}
\end{figure}

\begin{figure}
\includegraphics[width=0.45\textwidth]{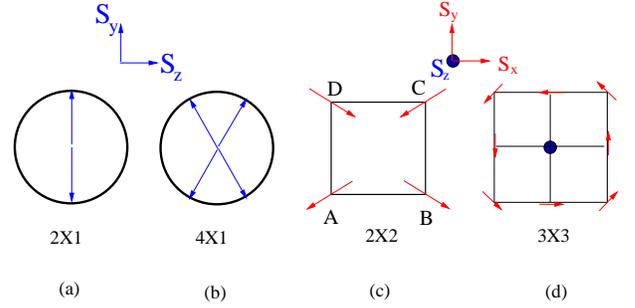}
\caption{
The Collinear, spiral, vortex and non-coplanar states in Fig.\ref{phasedia}.
(a) The $ 2 \times 1 $  ( Y-x  ) state $ S^{y}= (-1)^x $ (b)  The spin direction at the lattice sites
$ x=1,2,3,4 $ of the $ 4 \times 1 $ spiral state.
The inset shows the spin axis for the collinear and spiral states.
(c) The classically degenerate family of ( 2 in , 2 out )  $ 2 \times 2 $ vortex state.
(d) The $ 3 \times 3 $ non-coplanar skyrmion crystal ( SkX ) state with non-vanishing skyrmion density $ \vec{S}_i \cdot  \vec{S}_j \times \vec{S}_{k} \neq 0 $
happens near $ \alpha=\beta=\pi/3 $ which is the most frustrated regime in the Wilson loop \cite{rh}.
The inset shows the spin axis for the $ 2\times 2 $ vortex and  $ 3 \times 3 $  SkX states. }
\label{allphases}
\end{figure}

{\sl 2. The order from quantum disorders: selection of the quantum ground state: }
%{\sl  Order from disorder  near the $ \alpha=\beta=\pi/2 $ Abelian point. }
% The QFH model  along the dashed line  $ ( \alpha=\pi/2, \alpha < \beta ) $ have been throughly  studied in \cite{rh}.
% The $ 2 \times 1 $ ( $ Y-x $ state ) was proved to be the exact quantum ground state, so no quantum fluctuations at $ T=0 $.
It was shown the $ 2 \times 1 $ (  Y-x ) state\cite{rh,notation} is the exact quantum ground state along the anisotropic
 line  $ ( \alpha=\pi/2, \alpha < \beta ) $.
 Now we investigate the physics along the diagonal line $ \alpha=\beta $ near the Abelian point
 $ \alpha=\beta=\pi/2 $.
 At the classical level, the $ 2 \times 1 $  Y-x  state $ S^y=(-1)^x $ ( Fig.2a) is degenerate with the $ 1 \times 2 $  X-y  state $ S^x=(-1)^y $.
 In fact, due to a spurious $ U(1) $ symmetry,
 there is a family of states called $ 2 \times 2 $ vortex states in Fig.\ref{allphases}c: $ \mathbf{S}_i=
( (-1)^{i_y}\cos\phi,(-1)^{i_x}\sin \phi,0 ) $  which are  degenerate at the classical level.
% In general, this family breaks the $ [C_4 \times C_4]_D $  symmetry except at $ \phi= \pm \pi/4, \pm 3\pi/4 $.
% When $ \phi=0, \pi/2 $, it recovers the X-y and Y-x state respectively.
 The order from quantum disorder (OFQD)  mechanism  is needed to find the unique quantum ground state upto
 the  $ [C_4 \times C_4]_D $  symmetry in this regime.
% To perform a LSW calculation, one need to introduce a 4 sublattice structure $ A, B, C, D $ shown in Fig.\ref{allphases}.
 After making suitable rotations to align the spin quantization axis along the Z axis,
 we introduce  4 HP bosons $ a, b, c, d $ corresponding to the 4 sublattice structure
 $ A, B, C, D $ shown in Fig.\ref{allphases}c
 to perform a systematic $ 1/ S $ spin wave expansion \cite{swgap,sw1,japan} for a generic $ (\alpha, \beta) $:
 $ H=E_0+2JS\Big[H_2
	+\Big(\frac{1}{\sqrt{S}}\Big)H_3
	+\Big(\frac{1}{\sqrt{S}}\Big)^2H_4+\cdots\Big] $
 where $E_0=-2NJS^2(1-\cos2\alpha\sin^2\phi-\cos2\beta\cos^2\phi)$ is the classical ground state energy,
 $H_n$ denotes the $n$-th polynomial of the boson operators. $ H_2 $ can be diagonized by a unitary transformation,
 followed by a Bogoliubov transformation as:
\begin{align}
	H_2= E_2 + 2\sum_{n,k} \omega_n (k)\alpha_{n,k}^\dagger \alpha_{n,k}
\label{fourn}
\end{align}
    where $ n=1,2,3, 4 $ is the sum over the 4 branches ( due to the 4  sublattice $ A,B,C,D $ in Fig.\ref{allphases}c ) of spin wave spectrum in the Reduced BZ $ -\pi/2 < k_x, k_y < \pi/2 $ and
    $ E_2(\phi) =\sum_{k,n}[\omega_n(k)
	-(1-\cos2\alpha\sin^2\phi-\cos2\beta\cos^2\phi)/2] $ is the $ 1/S $ quantum correction to the ground-state energy.
%\begin{align}
%  $	E_2 =\sum_{k,n}[\omega_n(k)	-(1-\cos2\alpha\sin^2\phi-\cos2\beta\cos^2\phi)/2]  $.
%\label{quantumphi}
%\end{align}

   We first look at $ E_0 $ near the Abelian point $ \alpha=\beta=\pi/2 $.
   If $ \alpha > \beta $, it picks the Y-x state \cite{rh} with $ \phi=\pi/2 $.
   If $ \alpha < \beta $, it picks the X-y state with $ \phi= 0 $.
   Setting $ \alpha=\beta $, $ E_0 =-2NJS^2(1-\cos2\alpha) $ becomes $ \phi $ independent, indicating
   the classical degenerate family of states characterized by the angle $ \phi $ along the whole diagonal line $ \alpha=\beta $. Fortunately,
   the quantum correction $ E_2(\phi)=\sum_{k,n}[\omega_n(k,\phi)-\sin^2\alpha] $ does depend on $ \phi $.
As shown in Fig.3a, $E_2(\phi)$ reach its minimum at $\phi=0$ ( X-y  state ) or $\phi=\pi/2$ ( Y-x state) which is
related to each other by the $ [C_4 \times C_4]_D $ symmetry.
Expanding $E_2(\phi)$  around one of its minima  $\phi=0$:
\begin{equation}
E_2(\phi)= E_2^0 + \frac{1}{2} B \phi^2+ \kappa \phi^4 + \cdots
\label{B2}
\end{equation}
where one can identify the coefficient $ B(\alpha )$ plotted in the Fig.3b.
The OFQD selection of the Y-x or X-y state at $ \alpha=\beta $ shows that there is a direct first order transition from
the Y-x state to the X-y state, so at $ \alpha=\beta $, there is any mixture of the Y-x and X-y state  in Fig.1.
%Similar first order transition between vacancy induced supersolid ( SS-v ) and interstitial induced supersolid ( SS-i )
%and any mixtures of the two along the particle-hole symmetric line at the half filling in a triangular lattice were discussed in %\cite{dual1,dual2,dual3}.

%{\sl The excitation spectrum in the Y-x state. }
% However, we expect that the quantum fluctuations ( "order from disorder" ) encoded in the RFHM  will pick the
% $ 2 \times 2 $ vortex state as the ground state. Following the SWE developed in \cite{rh,rhh,rhht,rafh,sw1,sw2,sw3}, we will
% confirm this expectation and also determine the excitation spectra above this classical vortex state.
  Taking the Y-x state as the ground state, plugging $ \phi =\pi/2 $ into Eq.\ref{fourn},
  we find it supports the C$_\pi$ magnons \cite{rh,notation} at $ \mathbf{k}= (0,\pi) + \mathbf{q} $.
  They  condense along the diagonal line $  \arccos(1/\sqrt{6})\leq \alpha \leq \pi/2 $
  with the gapless relativistic dispersion:
\begin{equation}
	\omega_{-0}(q)=\sqrt{v_x^2 q_x^2+v_y^2 q_y^2}
\label{gapless}
\end{equation}
   where $ v_x=\cos(\alpha)/2,~
	v_y=\cos(\alpha)\sqrt{1-6\cos^2(\alpha)}/2 $. Obviously, both velocities vanish at the Abelian point $ \alpha=\beta=\pi/2 $
dictated by the hidden $ \tilde{\tilde{SU}}(2) $ symmetry. Moving away from the Abelian point, $ v_x$ keeps increasing,
but $ v_y$ increases first, reaches a maximum, then decreases, vanishes at
$ \alpha^{0}_{ic}= arccos(1/\sqrt{6} ) \sim 0.36614 \pi $, indicating a possible quantum Lifshitz transition (QLT).
As to be shown below, the gapless magnon mode in Eq.\ref{gapless} is just a spurious Goldstone mode due to the spontaneous
breaking of the spurious $ U(1) $ symmetry.
%As shown in the SM, the OFQD analysis generates a gap $ \Delta_B $ Eq.S2 and transfer it into a pseudo-Goldstone mode.
%However, the  quantum Lifshitz transition remains, but with a different dynamic exponent than
%$ ( z_x=1, z_y=2 ) $.

%{\sl  The magnon gap in the Y-x state generated by the order from disorder mechanism. }

%{\sl Quantum Lifshitz transition from the Y-x ( X-y ) to IC-SkX/Y-x ( IC-SkX/X-y ) state. }
%As shown in the last section, there is a gap $ \Delta_B $ opening at $ \vec{q}=0 $ along the diagonal line,
%so the  quantum Lifshitz transition point will shift to a smaller value of $ \alpha $.
% One can derive the  by incorporating the site dependence...it is not known how to incorporate the site dependence into
% the term generated from "order from disorder " mechanism...
%Obviously, the magnons always take the relativistic form Eqn.\ref{relagap} to any order of $1/S $, so
%it is justified to incorporate the gap $ \Delta_B $  into the spin-wave dispersion $\omega_k$ in Eqn.\ref{gapless} at the LSW order.
\begin{figure}[!htb]
	%\centering
\includegraphics[width=0.45\textwidth]{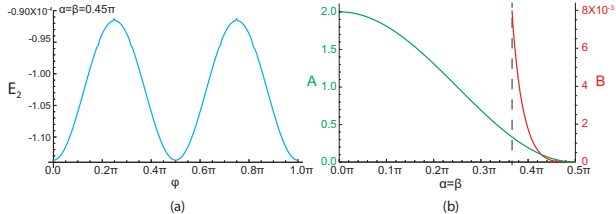}	
%\quad
%\includegraphics[width=0.2\textwidth]{KC.eps}	
	\caption{ The order from the quantum disorder (OFQD) and the gap opening on the spurious gapless mode in the Y-x state in Fig.1.
   (a) The quantum correction $ E_2 (\phi) $ to the ground-state energy
%	$\phi=0$ corresponds to X-y state and $\phi=\pi/2$ corresponds to Y-x state.
    picks up Y-x at $\phi=0$ or X-y at $\phi=\pi/2$ as the ground state which is related to each other
    by the $ [C_4 \times C_4]_D $ symmetry.
    (b) The classical coefficient $ A( \alpha)/J $ labeled by the left axis
    ( the green line on the left ) and the quantum one $ B(\alpha )/J $ labeled by the right axis
    ( the red line on the right ). Both vanish at
    the Abelian point $ \alpha=\beta =\pi/2 $  as $ \sim ( \pi/2- \alpha)^2 $ and are monotonically increasing function
    when moving away from the Abelian point. The Dashed line is located at
    $ \alpha^{0}_{in} \sim 0.3661 \pi $ where the Y-x state becomes unstable at the linear spin wave order.
    After incorporating the gap opening, the $ \alpha^{0}_{in} $ is shifted to a  smaller value $ \alpha_{in} \sim 0.3526 \pi $.
    The gap $ \Delta_B $ in Eq.\ref{gap}  keeps increasing
    when moving away from the Abelian point $ \alpha=\beta=\pi/2 $. }
\label{orderdis}
\end{figure}

{\sl 3. Order from quantum disorder (OFQD): the gap opening and the spectrum: }
  By using the spin coherent state path integral formulation \cite{aue,sachdev,swgap}, we will evaluate the gap at the minimum $ (0, \pi) $ of the
  $ C_{\pi} $ magnons in the  $ \tilde{\tilde{SU}}(2) $ basis \cite{rh}.
%   with $ \vec{\tilde{\tilde{S}}}_{i}  = R(\hat{x},\pi n_1)  R(\hat{y},\pi n_2) \vec{S}_{i} $.
  A general uniform state at $ \vec{q}=0 $ in the $ \tilde{\tilde{SU}}(2) $ basis can be taken as a
  Ferromagnetic (FM) state with the polar angle $ (\theta, \phi) $.
  After transforming back to the original basis by using  $ \tilde{\tilde{S}}_1=R_z(\pi)S_1,~
	\tilde{\tilde{S}}_2=R_y(\pi)S_2,
	\tilde{\tilde{S}}_3=R_x(\pi)S_3,
	\tilde{\tilde{S}}_4=S_4 $,  it leads to a  $ 2 \times 2 $ state characterized by the two angles $ \theta $ and $ \phi $.
%\begin{eqnarray}
%    S_1 & = & (-\cos\phi\sin\theta,-\sin\phi\sin\theta,\cos\theta), \nonumber   \\
%	S_2 & = & (-\cos\phi\sin\theta,\sin\phi\sin\theta,-\cos\theta),  \nonumber    \\
%	S_3 & = & (\cos\phi\sin\theta,-\sin\phi\sin\theta,-\cos\theta),  \nonumber    \\
%	S_4 & = & (\cos\phi\sin\theta,\sin\phi\sin\theta,\cos\theta)
%\label{22vortex}
%\end{eqnarray}
   Along the diagonal line, its classical energy becomes
$	H_0=J[-2 \sin^2 \alpha - 2\cos^2 \alpha \sin^2 \theta ] $
  which is, as expected, $ \phi $ in-dependent. Any deviation from the Abelian point picks up the XY plane
  with $ \theta=\pi/2 $. So it reduces to the $ 2\times 2 $ vortex state shown in Fig.2c.
  Expanding around the minimum $ H_0 =J[-2\sin^2\alpha+ 2\cos^2\alpha (\theta-\frac{\pi}{2})^2+\cdots] $ gives
  the stiffness $ A =2J\cos^2\alpha$ shown in Fig.\ref{orderdis}b. Using the spin coherent state analysis,
  we can write down the quantum spin action at $ \vec{q}=0 $:
\begin{equation}
  {\cal L}( \vec{q}=0 )= i S \cos \theta \partial_{\tau} \phi + \frac{1}{2} S^2 A ( \theta-\pi/2)^2 + \frac{1}{2} S B \phi^2
\label{action00}
\end{equation}
  where we put back the spin $ S $, the first term is the spin Berry phase term,
  $ A \sim (\pi/2-\alpha)^2 $ and $ B \sim (\pi/2-\alpha)^2 $ are from the  classical analysis
  and the OFQD analysis Eqn.4 respectively.
  Eqn.\ref{action00} leads to the gap
\begin{equation}
   \Delta_B= \sqrt{S A B} \propto \sqrt{S}
\label{gap}
\end{equation}
  which is beyond any $1/S $ expansion, so non-perturbative.
In fact, there are also corrections from the cubic $ H_3 $ and quartic $ H_4 $ terms in the spin wave expansion listed above Eqn.3,  but they only contribute to order of $ 1 $ which is subleading to the $ \sqrt{S} $ order in the $ 1/S $ expansion \cite{swgap,sw1,japan}.
  As shown in Fig.\ref{orderdis}b, both $ A $ and $ B $ are monotonically increasing along the diagonal line,
  so the gap also increase. Plugging their values at $ \alpha=\alpha^{0}_{in} = \arccos(1/\sqrt{6}) $,
  Taking $A/J=1/3, B/J \approx 8\times 10^{-3} $ and $S=1/2$,
  we find the maximum gap near the quantum Lifshitz transition $ \Delta_B/J \sim 0.036  $.

 In the SM1 \cite{SM}, we develop a new systematic non-perturbative scheme to evaluate not only the mass gap Eq.\ref{gap},
 but also the whole spectrum:
%Incorporating the gap $ \Delta_B $ into the spin-wave dispersion $\omega_k$ in Eqn.\ref{gapless} at the
%LSW order leads to \cite{confirm}:
%     whose evolution is shown in Fig.\ref{cictransition}
%    ( because $ B \ll A $, so we ignore its small contribution to $ \omega_q $ ).
%Because the spectrum along $ q_x $ is non-critical, so one can just put $ q_x=0 $:
 \begin{align}
	\omega_{-}(q_x, q_y)=\sqrt{ \Delta^2_B+ v_x^2 q_x^2 + v_y^2 q_y^2 + u^{2} q_y^4 + \cdots }
\label{gapadd}
\end{align}
    where $ v^{2}_y= a ( \alpha^{0}_{in} - \alpha ) $ changes sign at $ \alpha = \alpha^{0}_{in} $.
    From the gap vanishing condition \cite{loff} ( See also Eq.\ref{Yxp} ) at the IC- wave-vectors $ q_{ic}= \pm (\Delta_B/u)^{1/2} $, one can see
    the QLT is shifted to $ \alpha_{ic} = \alpha^{0}_{in}- 2 u \Delta_B/a  $.
    Plugging in the values of $ \Delta_B $ and $ u $,
    we find  $ q_{ic} \sim 0.18 \pi $. The shift is so small that $ \alpha_{ic} \sim 0.3526 \pi $
    remains larger than $\alpha_{33} \sim 0.3402 \pi $ ( to be defined in Sec.5 ) shown in Fig.\ref{phasedia}.
    So there must be an IC- phase intervening between the Y-x state and the  $ 3 \times 3 $ state
    when $ \alpha_{33} < \alpha < \alpha_{ic} $ in Fig.1.
%    In the following, we construct a GL action in terms of the pseudo-Goldstone mode $ \phi $ to describe the quantum Lifshitz transition.

{\sl 4. The Quantum Lifshitz transition (QLT) from the Y-x phase to IC-XY-y phase: }
 Here we construct an effective action in terms of the pseudo-Goldstone mode $ \phi $ to describe the quantum Lifshitz transition.
 This is a symmetry based phenomenological approach which is independent of the  $ 1/S $ expansion in the previous sections.
% However, the phenomenological parameters in the effective GL action can be evaluated by the $1/S $ spin-wave expansion
% which is a microscopic calculation. Of course, the microscopic calculation in the previous sections can guide us to
% construct the phenomenological effective GL action  consistent with all the symmetries of
% the microscopic Hamiltonian Eq.2
 Inside the Y-x phase along the diagonal line $ \alpha=\beta $,
 after integrating out the massive conjugate variable $ \theta- \pi/2 $,
% adding the effective potential Eq.\ref{B2} generated  from the OFQD mechanism,
 we reach the following effective GL action in the continuum limit consistent with all the symmetries of the microscopic Hamiltonian Eq.2
\begin{eqnarray}
{\cal L}_{Y-x}[ \phi ] & = & \frac{1}{2A} ( \partial_\tau \phi )^2  + v^2_x ( \partial_x \phi)^2
    + v^2_y ( \partial_y \phi)^2 + u^2 ( \partial^2_y \phi)^2  \nonumber  \\
    & + & \frac{1}{2} B \phi^2 + \kappa \phi^4 + \cdots
\label{Yx}
\end{eqnarray}

 In general, it is difficult to evaluate the values of the phenomenological parameters in Eq.\ref{Yx}.
 However, in the large $ S $ limit and away from the QLT point,
 they  can be evaluated by the microscopic calculations in the previous sections.
 Indeed, by contrasting Eq.\ref{Yx} with Eq.\ref{action00},\ref{gap},\ref{gapadd},
 one can see  $ A $ is from a classical contribution, $ B $ and $ \kappa $ are the effective potential Eq.\ref{B2} generated
 from the OFQD mechanism.
 Notably,  the coefficient $  v^2_y= a ( \alpha - \alpha^0_{in}  ) $ tuned by the SOC changes sign at $ \alpha= \alpha^0_{in} $. These matches between the microscopic calculations in a large $ S $ limit  and the  symmetry based effective action
 ensures the non-perturbative OFQD calculation in Sec.3 is indeed correct.

%  so  there is a quantum Lifshitz transition at the two IC- wavevectors at $ \pm q_{ic} $ where the gap vanishes.
 It is physically more transparent to re-write Eq.\ref{Yx} in the momentum space:
\begin{eqnarray}
{\cal L}[ \phi ]_{Y-x,D} & = &  \phi(- \omega_n, -q_x, -q_y)
[\omega^2_{n}/A +  v^2_x q^2_x + u^2 ( q^2_y - q^2_{ic} )^2          \nonumber  \\
 & +  & \Delta ] \phi( \omega_n, q_x, q_y) + \kappa \phi^4 + \cdots
\label{Yxp}
\end{eqnarray}
 where  $ \Delta= \Delta^2_B- \frac{a^2}{4u^2} ( \alpha-\alpha^0_{in} )^2 $
 is the tuning parameter of the QLT.

  The spin can be expressed in terms of the order parameter $ \phi $  when using the
  shift $ \phi \rightarrow \phi+ \pi/2 $ and setting $ \phi $ small.
\begin{equation}
  \mathbf{S}_i \sim (- (-1)^{i_y} \phi,(-1)^{i_x},0 )
\label{2times2order}
\end{equation}
%  where we assumed that the quantum fluctuations around the minimum $ \pi/2 $ is small, so the linear approximation is justified.

  So we conclude that when $ \Delta >0 $, $ \langle \phi \rangle =0  $, it is inside the Y-x phase.
  When $ \Delta < 0 $,  then
\begin{equation}
   \langle \phi \rangle = P_0 \cos ( q_{ic} y + \phi_0 )
\end{equation}
  where $ P_0, \phi_0 $ need to be fixed by the 4th order $ \kappa $ term.
  Substituting it into Eq.\ref{2times2order} shows that the system is in the IC-XY-y phase \cite{notation}.
  The smallness of  $ \langle \phi \rangle $ justifies the expansion in Eq.\ref{B2}.
  The transition from the Y-x to the IC- state is a quantum Lifshitz transition with the dynamic exponent $ z=1 $.
  All the quantum critical scalings will be evaluated in \cite{unlong}
  by $ 1/N $ expansion and $ 4 - \epsilon $ expansion with $ \epsilon=1 $.

{\sl 5. The $3 \times 3 $ non-coplanar SkX phase: }
Near $ \alpha=\beta=\pi/3 $, it is natural to take a $3\times3$ ansatz: $S_{(i_x,i_y)}=S_{(i_x+3m,i_y+3n)}$ with $m,n\in \mathbb{Z}$.
We  estimate its classical ground-state energy by minimizing
$ E_{3\times3}(\{\phi_i,\theta_i\}_{0\leq i\leq 9})$ over its 18 variables.
Along  the diagonal line ($\alpha=\beta$), as long as $\alpha$ is not too small, the minimization of $E_{3\times3}$
always leads to the $ 3 \times 3 $ SkX state which respects the $ [C_4 \times C_4]_D $ symmetry ( Fig.\ref{allphases}d ).
The total spin in the $ 3 \times 3 $ unit cell is $	S_{\rm unit}=\sum_i S_i=(0,0,4 \times 10^{-3}) $
which has exact vanishing $ S_x, S_y $ components, but still a small non-vanishing $ S_z $ component.
%This is in sharp contrast to the case near $ \alpha=\beta=\pi/2 $ where the classical analysis only leads to the degenerate
%family of $ 2 \times 2 $ vortex states shown in Fig.\ref{allphases}.
%A quantum " order from disorder " analysis is needed to show the $ 2 \times 2 $ vortex state phase separates into
%any mixtures of the Y-x state and X-y state along the diagnose line.

Comparing the classical ground energy of the $ 3 \times 3 $ SkX with that of the  Y-x state
$E_{Y-x}=-2J\sin^2\alpha$ leads to a putative first order transition between the two states at
$\alpha_{33} \approx0.340188\pi $ which is smaller than  $ \alpha_{ic} \sim 0.3526 \pi $ ( Fig.1 ).
So a putative direct first order transition between the Y-x state and the $ 3 \times 3 $ SkX splits into
two second order QLTs with $ z=1 $ with the IC-XY-y phase intervening between them in Fig.1.
%In fact, $ \alpha_{33} $ also shifts to a smaller value due to the intervening of the IC-SkX/Y-x phase, but
%for simplicity, we still use the same symbol.  The point $  \alpha= \alpha_{33} $ in Fig.\ref{phasedia} is a bi-critical point.
%Similarly, by using the LSW, we can determine the excitations spectra above the $ 3 \times 3 $ SkX.
%We expect the transition from the $ 3 \times 3 $ SkX to the IC-XY-y state  is also a 2nd order quantum Lifshitz transition.
When approaching $ \alpha=\beta $ from the anisotropic line $ ( \alpha=\pi/2, \beta ) $ from the right \cite{unlong}, we find
$  \alpha= \alpha_{33} $  lies on the constant contour line of the C-IC magnons $ (0, k^{0}_y ) $  at $ k^{0}_{y} \sim \pi-0.24 \pi $.
So $ 0.18 \pi < q^{0}_{y} < 0.24 \pi  $ in the IC-XY-y phase $  \alpha_{33} < \alpha < \alpha_{in} $ (Fig.1).

{\sl 6. Possible experimental implications:}
 %The heating issue has been well under control in the weak coupling limit in recent cold atom experiments \cite{expk40,expk40zeeman,2dsocbec}..
 %However, the heating issue gets more serious as the coupling increases.
 %Many new techniques \cite{clock,clock1,clock2,SDRb,ben} are being developed to control the heating effects
 %even in the strong coupling limit.
% The RFHM Eq.\ref{rhgeneral} can only be reached in the strong coupling limit.
 The heating issue has been well under control in the weak coupling limit in recent cold atom experiments
 \cite{expk40,expk40zeeman,2dsocbec,clock,clock1,clock2,SDRb,ben,gong}. So various exotic magnetic  superfluid phenomena
 can be observed in the current cold atom experiments.
 However, it gets worse as the coupling increases.
 The RFHM Eq.\ref{rhgeneral} can only be reached in the strong coupling limit.
 So the rich magnetic Mott phenomena discovered in this manuscript can be observed only after the heating issue
 can be resolved in the strong coupling limit.
 Now, we turn its qualitative applications in the strongly correlated 4d
 or 5d materials with strong SOC.

% As mentioned in the introduction, the rich magnetic phenomena discovered in this manuscript can be observed in
% future cold atom experiments as soon as the heating problem can be overcame in the strong coupling limit\cite{clock,clock1,clock2,SDRb,ben}.
% Here we turn its qualitative implications in the strongly correlated 4d or 5d materials with strong SOC.
 Naively, due to its microscopic bosonic nature, the RFHM Eq.\ref{rhgeneral}  may not be useful to describe the magnetism in
 various materials with SOC. However, the RFHM can be expanded  \cite{rh} as Heisenberg-Kitaev (or Compass)-
 Dzyaloshinskii-Moriya (DM) \cite{dm1} form:
\begin{equation}
 H_{R}  =  \sum_{\langle i j \rangle  } J_{H} \vec{S}_{i} \cdot \vec{S}_{j} +
     \sum_{\langle i j \rangle a } J_{K} S^{a}_{i} S^{a}_{j}  + \sum_{\langle i j \rangle a } J_{D} \hat{a} \cdot \vec{S}_{i} \times \vec{S}_{j}
\label{expand}
\end{equation}
 where $ \hat{a}=\hat{x}, \hat{y} $ and $  J_{H}= \cos 2 \alpha, J_{K}= 2 \sin^{2} \alpha, J_{D}=  \sin 2\alpha $.
 One can estimate their separate numerical values near the in-commensurate phase ( IC-XY-y )
 $ \alpha= \alpha^{0}_{in}=\arccos \frac{1}{\sqrt{6}} $ in Fig.\ref{phasedia}:
the Heisenberg term $ J_{H} \sim -2/3 $ is AFM,
the Kitaev term  $ J_{K} \sim 5/3 $ is FM, the DM term $  J_{D} \sim \sqrt{5}/3 $.
So the model becomes a dominant FM Kitaev term plus a small AFM Heisenberg term and a small DM term.
This is indeed the case  in the so called  5d Kitaev materials such as $ A_2 Ir O_3 $ with $ A=Na_2, Li_2 $ or more recent 4d materials $ \alpha-Ru Cl_3 $. So far,
only a Zig-Zag phase or an IC- phase were observed experimentally \cite{kitaevlattice,kitaevlattice1},
no quantum spin liquids \cite{kit,kit123} have been found.

{\sl 7. Discussions: }
It is instructive to  contrast the Quantum phenomena achieved here by the  analytic perturbative and non-perturbative methods
with those results achieved by classical Monte-Carlo simulations in the two earlier works \cite{classdm1,classdm2}.
%We compare the results achieved in this manuscript with the two earlier works
%\cite{classdm1,classdm2}.
The authors in \cite{classdm1,classdm2} did {\sl classical} Monte-Carlo {\sl simulations} using the
representation Eq.\ref{expand} on a small finite size system.
These two numerical papers did not have the concepts of the frustrations due to the Rashba SOC.
Ref.\cite{classdm1} found the classical $ 2 \times 1 $, $ 3 \times 3 $ SkX and $ 4 \times 1 $ states in Fig.1.
It also found a Ferromagnetic (FM) state near the orgigin $ \alpha=\beta =0 $.
Ref.\cite{classdm2} found the classical $ 2 \times 2 $ vortex,  $ 3 \times 3 $ SkX and $ 4 \times 1 $ states in Fig.1.
Our work study the {\sl  quantum effects} on  the RFHM Eq.2 {\sl analytically }.
In Sec.2, we found  the $ 2 \times 2 $
vortex is classically degenerate with the Y-x and X-y state, but
the " order form quantum disorder" (OFQD)  mechanism  picks up either Y-x and X-y state as the quantum ground state.
In Sec.3, we also evaluated the excitation spectrums corrected by the mechanism.
This anlysis also leads to the instability of the Y-x
( or X-y ) state to the IC-SkX phase.
In Sec.4, we constructed an effective action to describe the quantum Lifshitz transition (QLT) with the dynamic exponent $ z=1 $ in Fig.1
and also identify the spin-orbital structure of the IC-SkX phase.
Of course, it would be impossible to detect the quantum IC-SkX phase  by
any classical  Monte-Carlo simualations at any finite size system, let alone to study the QLT.
Only by the controlled, non-perturabative  analytical calculations, one can show there must be an in-commensurate phase intervening between
the collinear Y-x phase and the non-coplanar $ 3 \times 3 $ SkX phase.
Of course, the quantum model Eq.2 presents very serious sign problem to quantum Monte-Carlo simulation.
So the classical {\sl classical} Monte-Carlo {\sl simualations} used in \cite{classdm1,classdm2} can not be extended to
study the novel quantum and topological phenomena address in this work.

As alerted above, the second term in Eq.\ref{expand} is a quantum compass model in a square lattice
instead of the Kitaev model in a honeycomb lattice. In order to have quantitative impacts on the 3d or 4d Kitaev materials \cite{kitaevlattice,kitaevlattice1},
it is important to extend the results achieved here in a square lattice to
a honeycomb lattice with 3 SOC parameters $ \alpha,\beta,\gamma $.

We thank Wei Ku for the hospitality during our visit at T D Lee institute.
We acknowledge AFOSR FA9550-16-1-0412 for supports.


\begin{thebibliography}{99}
\bibitem{notation} 
    Here we still use the same notation used in \cite{rh}. In the Y-$(\pi,0) $ called Y-x state, the first letter indicates
    the spin polarization, the second letter indicates the $(\pi,0) $ orbital order.
    Along the anisotropic line $ (\alpha=\pi/2, \beta) $, it supports three kinds of magnons C$_{0}$, IC and C$_{\pi}$ with
    their minimum at $ (0,0), (0, \pm k^{0}_y ) $ and $ (0,\pi) $ respectively.
    The IC-XY-y phase means IC- in the spin XY plane with the IC- momentum along the y axis.
\end{thebibliography}
\end{document}